# Multi-horizon solar radiation forecastingfor Mediterranean locations using time series models

Cyril Voyant[1,2*], Christophe Paoli[1],Marc Muselli[1], Marie-Laure Nivet[1]

[1] University of Corsica, CNRS UMR SPE 6134, 20250 Corte, France

[2]Castelluccio Hospital, Radiotherapy Unit, BP 85, 20177 Ajaccio, France

*Corresponding author, tel/fax +33(0)4 952 936 66/937 97, voyant@univ-corse.fr

Abstract.

Considering the grid manager's point of view,needsin terms ofprediction of intermittent energy like thephotovoltaic resourcecan be distinguishedaccording to theconsideredhorizon: following days (d+1, d+2 and d+3), next day by hourly step (h+24), next hour (h+1) and next few minutes (m+5 e.g.). Through this work, we haveidentified methodologies using time series modelsfor thepredictionhorizonof global radiationand photovoltaic power. What wepresent here isa comparison of differentpredictorsdeveloped and testedtoproposea hierarchy.For horizons d+1 and h+1, without advanced ad hoc time series pre-processing (stationarity) we find it is not easy to differentiate between autoregressive moving average (*ARMA*) and multilayer perceptron (*MLP*). However we observed that using exogenous variables improves significantly the results for *MLP* . We have shown that the MLP were more adapted for horizons h+24 and m+5. In summary, our results are complementary and improve the existing prediction techniques with innovative tools: stationarity, numerical weather prediction combination, *MLP* and *ARMA* hybridization, multivariate analysis, time index, etc.

# 1. Introduction

There are lots of alternatives to greenhouse gas emissions generated by fuels combustion [1,2]. It is particularly the case ofphotovoltaic (*PV)* and wind energy sources, which one of the main advantages is the renewable and inexhaustible aspects and the main disadvantages are related to their intermittencies. Thisvariability is related to winter/summer transition, to day/night transition and to the opacity of atmosphere [3,4]. To overcome these problems, which can be prohibitive, three solutions can be envisaged: split and better distribute the total available power, predict the resource to manage the transition between different energies sources and store the energy excess to redistribute it at the right time [5,6]. This paper deals only with the second solution: the forecasting of the renewable energy sources. The optimizationand the management of energy system are really a challenging issue especially when there are insufficient renewable energies to meet the demand. It is essential to anticipate the global radiation decrease (or increase) for an ideal transition. Several methods have been developed by experts around the world and can be divided in two main groups: (i) methods using mathematical formalism of Times Series (*TS*), (ii) numerical weather prediction (*NWP*) model and weather satellite imagery. The technique used depends on considered source, and on their startup delay (from five minutes to 1 hour). Note that for an ideal management it is appreciable to know the eventual fluctuations one or two days ahead. These temporal characteristics define the horizon of the prediction to consider. According to thehorizonsome of these methodsare more effectivecomparedto others [8]. Considering the grid manager's point of view, needsin terms ofprediction can be distinguishedaccording to theconsideredhorizon: the resourcethat will be availableon the following days (d+1, d+2 et d+3), the next day by hourly step (h+24), during the next hour (h+1), and in the five next minutes (m+5). Thesehorizonsallow understandingthe various aspects ofthe prediction: the medium term, the short term and the very short term. The d+1and d+2 predictionsare importantfor the managerbecause theyhaveimmediate industrial applications and economic impacts especially in the case of small and relatively isolated electric grids. Indeed,in thiscase it is essential to organize and anticipate the fossil stocks. Concerning the h+1 horizon, itcorresponds more or less to theignition delayof the thermalsystem. In fact, starting a heat engine takesabout 30 minutes; the manager mustbeable to predict theintermittent energycutsat least 1 hourin advance. Concerning the h+24 prediction, itsinterestcombinesthe two precedents. The knowledge24 hours inadvance of therenewable energyenables betterinventory management concerning fossil fuel, and an anticipationofthecritical moments wherethe grid managermust bevigilant. Finally, thefewminutes horizonconcern for example the means of production related to hydroelectric power plantsand to gas turbines. Indeed, just a few minutes are necessary toelectricity to beavailable in these cases.Wecan also note that short term forecasting (now-casting) can be very useful to control indoor climate in buildings with automation system. Thus it seemedinteresting to comparedifferentmethods based upon the analysis of historical *TS* of global radiation for several horizons: d+1, h+1, h+24 and m+5. In this paper we propose, horizon by horizon, a classification of predictor tested on various Mediterranean towns. Our goal is to provide robust predictors with the most generic approach possible.

In the next, the time series forecasting models proposed in the literature are first reviewed. In section 3 we will detail the methodologies of prediction we have tested, taking care to explain the *TS* formalism dedicated to the global solar radiation modeling and the need to make it stationary (time series pre-processing). Then we will expose the result of comparison between modeling and measure in the daily case, hourly case and five minutes case. Finally we will close the paper with a comparison of the results against those of the literature, emphasizing the link between predictor performance and type of horizon.

## 2. Review on time series forecasting models

In this section, we present a review of the literature on time series forecasting models for global radiation. Optimal use of renewable energy requires a good characterization and good predictive potential for size detectors or estimate the potential energy power plants [9,10]. There are a lot of models allowing *TS* predictions. It is possible to list them into four groups [11,12]:

- naive models are essential to verify the relevance of complex models. Include persistence, average or the k-nearest neighbors (k-NN) [13-16];

- conditional probability models are rarely mentioned in the literature regarding global radiation. Include Markov chains and predictions based on Bayesian inference [17-22];

- reference models based on the family of autoregressive moving average, *ARMA* [23,24];

- connectionist models (artificial neural network) and more particularly the Multi-Layer Perceptron (*MLP*) which is the artificial neural networks architecture the most often used [25-27].

The following deals with the two last groups: *ARMA* and neural network models. Indeed *ARMA* is the most classical and popular for time series modeling and artificial neural network seems to be the best alternative to conventional approaches. As climate of the earth is dominated by non-linear processes, *ANN* by its non-linear nature is effective to predict cloudy days and so solar radiation. Concerning the prediction of solar radiation, we can cite works of Mellit [26,27] in which it is possible to find a synthesis of the coupling of *MLP* with global radiation. In addition to these works, there are others related to the prediction of weather data such as solar radiation [28-35]. Neural networks have been studied on many sites and researchers have shown the ability of these techniques to accurately predict the time series of meteorological data [32]. Table 1 presents several representative examples of the use of artificial neural networks (*ANN*) methods applied to the modeling or prediction of solar radiation and *PV* energy in the 2000s. For the years prior to 2000, the interested reader may also refer to the article Mellit [26]. For all the articles presented in Table 1 we can see that the errors associated with predictions (monthly, daily, hourly and minute) are between 5% and 10%. However we see that the *MLP* can be used with exogenous parameters or coupled with other predictors (Markov, Wavelet, etc.). In the Mellit and Kalogirou article review [26], we find that 79% of Artificial Intelligence (*AI*) methods used in weather prediction data are based on a connectionist approach (*ANN*). We can also cite the use of fuzzy logic (5%), Adaptive neuro fuzzy inference system (*ANFIS*) (5%), networks coupling wavelet decomposition and *ANN* (8%) and mix *ANN*/Markov chain (3%). In sum, the use of *ANN*, especially the *MLP* represents a large majority of research works. This is the most commonly used technique. Other methods are used only sporadically.

| authors | topic | location | horizon | error | conclusions |
|---|---|---|---|---|---|
| [58] Almonacid(2010) | Estimation of PV energy | Spain (Jaèn) | monthly | MAPE = 7.3 % | MLP better than reference models (bilinear interpolation method and Blaesser's method) |
| [31] Behrang et al. (2010) | Global radiation modeling with different ANN | Iran (Dezful) | d+1 | MAPE = 5.2 % | MLP with exogenous inputs is very efficient (8 models compared) |
| [56] Benghanem and mellit(2010) | Global radiation modeling with avec RBF.MLP and standard regression | Saudi Arabia (Al-madinah) | d+1 | $R^2$=0.98 | RBF is the most efficient, moreover the approach is validated on PV system (8 models are compared) |
| [27] Mellit et Pavan(2010) | Global radiation forecasting at horizon with ANN | Italy (Trieste) | h+24 | $R^2$>94 % | MLP validated on PV wall (no other compared predictors) |
| [59] Azadeh et al. (2009) | Global radiation modeling with ANN | Iran (6 cities) | monthly | Accuracy = 94 % (error = 6 %) | MLP better than Angström model |
| [57] Chaabene and Ben Ammar(2008) | Global radiation prediction with hybrid MLP with fuzzy logic, ARMA and Kalman filters | Tunisia (Energy and Thermal Research Centre) | d+1 m+5 | nMBE= -9.11 % nRMSE< 10 % | Dynamic predictions are considered coupling ARMA, Kalman filter and neuro-fuzzy estimators |
| [60] Jiang (2008) | Diffuse radiation prediction with MLP | China (8 cities) | monthly | Accuracy = 95 % | The methodology is validated on the entire Chinese territory (compared to two empirical models) |
| [52] Mubiru and Banda (2008) | Global radiation modeling with different MLP | Uganda (4 sites) | d+1 | RMSE = 107 Wh/m² | MLP better than 5 empirical models |
| [53] Bosch et al. (2008) | Global radiation modeling | Spain (13 sites) | d+1 | nRMSE = 7.5 % | The MLP can be used in the mountainous area. the error is acceptable (no comparison with other methods) |
| [55] Elminir et al. (2007) | Prediction of diffuse radiation with MLP | Egypt (3 stations) | h+1 d+1 | Standard error = 4.2 % Standard error = 9 % | MLP better than 2 linear regressions models |
| [61] Mellit et al. (2006) | Prediction of global radiation with MLP and wavelets | Algeria (36°43′ N; 3°2′ E) | d+1 | MAPE< 6 % | Method validated for the PV output and various meteorological data. More than 7 models are compared (AR, ARMA, MTM, MLP, RBFN, Wavelet networks, etc.) |
| [62] Cao and Cao (2005) | Prediction of global radiation with recurrent MLP and wavelets | China (Shanghai) | d+1 | nRMSE = 8 % (with wavelet) and 35 % without wavelet | Wavelet decomposition improves the prediction |
| [63] Mellit et al. (2005) | Global radiation modeling with MLP and Markov approach | Algeria (4 sites) | d+1 | nRMSE = 8 % | MLP better than AR,ARMA and Markov chains |
| [64] Sozen et al. (2004) | Global radiation modeling with MLP | Turkey (17 stations) | d+1 | MAPE< 7 % | Training and test areas are relocated, MLP is robust. The comparison is done with classical regression models |
| [65] Reddy and Manish (2003) | Global radiation modeling with MLP | India (2 stations) | h+1 | MAPE = 4 % | MLP better than 3 classical regression models |
| [36] Sfetsos andCoonick (2000) | Global radiation forecasting with MLP | Corsica (41.55°N, 8.48°E) | h+1 | RMSE = 27.6 W/m² | Multivariate MLP modeling improves the prediction. 13 Models are tested (ARMA, RNFN, ANFIS, etc.) |

Table 1: representative examples of the use of ANNs methodappliedto the modeling or prediction of solar radiation and PV energy from 2000s

Also in this literature review [26], the results of different researches considering a lot of places, were compared. The prediction error (MAPE in this case) of monthly global radiation induced by the use of an *ANN* is estimated between 0.2% and 10.1% depending on the city and the architecture considered (median= 4%). The results presented are so disparate they seem incomparable. However, we must consider that in some locations the cloud occurrences are minimal while others are subject to much less forgiving climates. Concerning the global radiation, Sfetsos [36] has showed that neural networks generated an error of 7% and *ARMA* methodologies, an error of 8%. Behrang et al. [37] have compiled a list of the predictions error with neural networks for global radiation. For identical locations, the errors can double or even triple. The conclusions on the *MLP* can be generalized to other predictors. According to the literature, the parameters that influence the prediction are manifold, so it is difficult to use the results from other studies. Considering this fact, it may be interesting to test methods or parameters even though they have not necessarily been proven in other studies. Based on the foregoing, all parameters inherent to the *MLP* or *ARMA* method must be studied for each tested site.

After literature review and considering the difficulty to make definite conclusion we wanted to study estimators which are little or very rarely studied in the renewable energy field. Thus, we tried a prediction methodology based on Bayesian inferences. There are many works on the coupling with other predictors such as neural networks [38,39] or as discriminant test for variables selection [40]. However, this technique is widely used in econometrics, through very theoretical publications cannot really compare with other prediction methods. We can especially mention XiangFei [41], which showed that the Bayesian inferences allow an estimate equal to autoregressive (*AR*) model with non-stationary variables. The error in the studied series is close to 10% for both models. Concerning Markov chains, they are rarely used in energy, according to the paper of Hoacaoglu [42] there is a prediction error of 6% for daily radiation and for Muselli et al. [43] an error on the *PV* predicted energy on horizontal surface equal to 10%. Based on these results, we chose to incorporate this type of predictor in our study. The other three studied estimators are persistence, k-NN and average which are easy to implement. Indeed, there is no learning phase, and few constraints are needed to use them (stationarity, pretreatment, assumptions, etc.). Although advanced methods provide better results, we think it is important to keep in mind the balance between model complexity and quality of prediction. For this reason, it is necessary to compare the sophisticated models against "naïve" models [4,15,44,45]. According to the references listed above, the following remarks can be made:

- *ANN* and *ARMA* models seem to be the most popular time series predictors;
- it is very difficult to compare or evaluate predictors because many of them looks like to be site and horizon dependant;

- there is no convention dealing with errors estimation tools (e.g. seasonal errors best for certain days), neither than with data test selection.

Considering these limitations we propose for each considered horizon a homogeneous experimental protocol.

## 3. Materials and methods

The methodology used in this work is based on time series forecasting. A Time Series (*TS*) is intuitively defined as an ordered sequence of past values of the variable that we are trying to predict [24]. Thus, the current value at *t* of the *TS* x is noted $x_t$ where *t*, the time index, is between 1 and *n*, with *n* is the total number of observations. We call *h* the number of values to predict. The prediction of time series from (*n+1*) to (*n+h*), knowing the historic from $x_1$ to $x_n$, is called the prediction horizon (horizon 1,…, horizon h). For the horizon 1 (the simplest case), the general formalism of the prediction will be represented by Equation 1 where $\epsilon$ represents the error between the prediction and the measurement, $f_n$ the model to estimate and *t* time index taking the (*n-p*) following values: *n, n-1,…, p+1, p*. Where n is the number of observations and p the number of model parameters (it is assumed that n ≫ p).[44,45]

$$x_{t+1} = f_n(x_t, x_{t-1}, \ldots, x_{t-p+1}) + \epsilon(t+1) \qquad \text{Equation 1}$$

Studies in finance and econometrics have yielded many models more or less sophisticated. Some of these models have been applied in the case of the prediction of global solar radiation. To estimate the $f_n$ model, a stationarity hypothesis is often necessary. This result originally shown for *ARMA* methods [23,24] can be also applicable for the study and prediction with neural network [46,47]. We can also note that few authors suggest that periodic nature of a time series can also be captured from the *AI* models like *MLP*, very often with the inclusion of a time indicator [36]. However we have considered that in practice, the input data must be stationary to use an *MLP*. In previous works [44,45], we have developed sophisticated methods to make the global radiation time series stationary. We have demonstrated that the use of the clear sky index (*CSI*) obtained with Solis model [48] is the more reliable in Mediterranean places. As the seasonality is often not completely erased after this operation, we use a method of seasonal adjustments (seasonal variance corrected by periodic coefficients) based on the moving average [24] (*CSI*$^*$). The chosen method is essentially interesting for the case of a deterministic nature of the series seasonality (true for the global radiation series) but not for the stochastic seasonality [23]. It is also possible to use a variant of *CSI*, considering only the radiation outside the atmosphere, we obtain in this way the clearness index (k) [49] and k$^*$ with the previous method of seasonal corrections.

Considering the limitations described at the end of the section 2, we decided to established a homogeneous experimental protocol for each considered horizon. Thereby, for all horizons studied (d+1, h+1, h+24 and m+5), we have compared *ARMA* and *MLP* predictors against at least one naive predictor (e.g. persistence). We focused our work on a general methodology for estimating the prediction error:

- test of prediction over a long period, not on "well chosen" days;

- use of *RMSE* to penalize large deviations [50];

- normalization of *RMSE* for comparisons on many sites:
$$nRMSE = \sqrt{E[(\hat{x} - x)^2]/\langle x^2 \rangle} \qquad \text{Equation 2}$$

- no cumulative predictions except for specific studies which has the effect of average the error and decrease it;

- distribution of errors according to seasons because the energy consumption is not the same throughout the year;

- tests on several locations, in order to avoid phenomena regional climates;

- use of a naive predictor as a reference for prediction to evaluate the proposed methodology (balance between model complexity and quality of prediction);

- use of confidence interval to define margin of error, as e.g. the classical IC95%, in order to provide information on the prediction robustness.

For *ARMA* and *MLP* methods, we have studied the impact of stationary process for the indexes *CSI*, k and relative seasonal adjustments (*CSI\** and k*). Concerning *MLP*, we studied the contribution of exogenous meteorological data (multivariate method) at different time lags and data issued from a numerical weather prediction model (*NWP*). The confidence interval has been calculated after at least six training simulations. We also studied the performance of a hybrid *ARMA*/*ANN* model from a rule based on the analysis of hourly data series. Finally we evaluated for each method the error estimation for annual and seasonal periods: Winter, Spring, Summer and Autumn. It should be noted that due to the difficulty to obtain data, the protocol could not be followed homogeneously for all data. The following section presents the results and for each horizon in chronological order.

## 4. Results

Data used in the experiments are related to the French meteorological organization database. As manipulations on horizons proceed, this database was expanded iteratively. Our goal is to provide robust and predictive methodology as generic as possible, avoiding falling into the specifics of a place. The non-homogeneity strict of manipulations is due to this typical construction. In fact it is very difficult to obtain quality data. At the beginning there was not much data available and after first experiments it seemed to be interesting to test our method on a larger sample. The table below lists for each horizon all manipulations performed and the data associated.

| Horizon | Manipulations performed | | | Data associated |
|---|---|---|---|---|
| | Predictor used | Stationary method | Variable selection | |
| d+1 | Mean, persistence, *SARIMA*, Bayesian inference, Markov chains, k-NN, *ANN* | *CSI*, k, *CSI\**, k\* | *PACF*, cross correlation | Ajaccio (1971:1989) and Bastia/Ajaccio (1998:2007) |
| h+1 | Mean, persistence, *ARIMA*, *ANN* | *CSI*, k | *PACF*, cross correlation, linear regression | Ajaccio/ Bastia/ Marseille/ Montpellier/ Nice (1998:2007) |
| h+24 | Persistence, *ARMA*, *ANN* | *CSI*, k | *PACF*, cross correlation | Ajaccio (1999:2008) |
| m+5 | Persistence, *ARMA*, *ANN* | *CSI*, k | *PACF*, cross correlation | Ajaccio (2009,2010) |

Table 2: list of manipulations performed and data associated with each horizon.

For the most complete horizon (hourly case), the data used to test models are from 5 coastal cities located in the Mediterranean area and near mountains: Montpellier (43°4'N / 3°5'E, 2 m alt), Nice (43°4'N / 7°1'E, 2 m alt), Marseille (43°2'N / 5°2'E, 5 m alt), Bastia (42°3'N / 9°3'E , 10 m alt) and Ajaccio (41°5'N / 8°5'E, 4 m alt). The available data are global radiation, pressure (*P*, Pa; average and daily gradient[1], measured by numerical barometer during 1 hour), nebulosity (*N*, Octas), ambient temperature (*T*, °C; maximum, minimum, average and night[2], measured done during an half hour), wind speed (*Ws*, m/s; average at 10 meters, measured during the 10 last minutes of the half hourly step), peak wind speed (*PKW*, m/s; maximum speed of wind at 10 meters, measured during 30 minutes), wind direction (*Wd*, deg at 10 meters measured during an half hour), sunshine duration (*Su*, h, computed with the global radiation series and the power threshold 120 W/m²), relative humidity (*RH*, % instantaneous measure at the end of the half-hour) and rain precipitations (*RP*, mm, 5 cumulative measures of 6 minutes during the half-hour). The data are transposed into hourly values by Météo-France organization.

## 4.1. Daily case

---

[1] Difference between the mean pressure of day *j* and day *j-1*
[2] Measured at 3:00 AM

As the knowledge of the available solar energy for the next days allows fossil energy provision and interconnection energy management, daily horizon is very important. For this horizon and for all studied models, the years 1971-1987 are the basis of learning and the two years from 1988 to 1989 are dedicated to the test of the prediction. With this horizon, the method based on average, Markov chains, k-NN and Bayesian inferences are tested. For all this methodology the results are equivalent, the error (*nRMSE*) is close to 25.5% (from 25.1 for Markov chains to 26.13 for the persistence). Without stationarization and exogenous inputs, the two predictors *ARMA* and *MLP* are more efficient than other methods; the errors of prediction are smaller than 22% and relatively close. The *MLP* is noted as: *(Endo$^{Ne}$)xN$_h$x1* where $N_e$ is the number of endogenous nodes and $N_h$ the number of hidden neurons. For this first study, where only endogenous data are considered, these two predictors are equivalent and outperform other approaches. If now we make the *TS* stationary by using k or *CSI* and seasonal adjustments (k$^*$ and *CSI$^*$*) we note that the error of prediction decreases. The best results are related to the k$^*$ and *CSI$^*$* pretreatments and are shown in the Table 3. With these methodologies the errors are reduced by 1.5 points (*nRMSE* =20.2% for k$^*$ and *nRMSE* =20.3% for *CSI$^*$*). Indeed, it is necessary to adapt the models and architectures to the new dynamics of the signal. The optimization leads to use the model *ARMA*(2,2), while for the *MLP* configuration remains unchanged.

|  | **Raw data** | **Statio k$^*$** | **Statio CSI$^*$** |
|---|---|---|---|
| *ARMA* | *21.18 ± 0%*<br>*AR(8)* | 20.31 ± 0%<br>*ARMA(2.2)* | 20.32 ± 0%<br>*ARMA(2.2)* |
| *PMC* | 20.97 ±0.15%<br>*Endo$^{1-8}$x3x1* | 20.17 ± 0.1%<br>*Endo$^{1-8}$x3x1* | 20.25 ± 0.1%<br>*Endo$^{1-8}$x3x1* |

Table 3: prediction errors for *ARMA* and *MLP* (*nRMSE* ± IC95%). Predictions done for years 1988 and 1989.

For more details on results of other methods (persistence, Bayesian, KNN, etc.), the reader can refer to our previous work [15,44]. Again, the *MLP* and *ARMA* methods appear to be equivalent for d+1. Indeed, with or without the use of seasonal adjustments, it is very difficult to prioritize them. It seems, in the particular case that we just examined, that *MLP* based results are also convincing than *ARMA* based results. Regarding the comparison between the two stationary methodologies (k$^*$ and *CSI$^*$*), it is not possible to conclude, averages are not significantly different. However, make stationary the *TS* improves the prediction error both for *ARMA* and *MLP*.

Once finished these first experimentations, we decided to explore the multivariate option. In order to increase the confidence degree of our conclusions we choose to make our test considering two locations: Ajaccio and Bastia (where forecasting is considered to be more difficult). Indeed one of the particularities of the *MLP* use is based on the possibility to do multivariate regressions. The use of the exogenous data should better model the phenomena. The *MLP* is noted as: *(Endo$^{Ne}$E$^{Me}$)xN$_h$x1* where *Ne* and *Me* are the numbers of endogenous and exogenous nodes. For Ajaccio for example the better model of *MLP* with exogenous data is *(Endo$^2$Su$^1$N$^1$)x3x1* while for Bastia it is *(Endo$^4$Su$^1$RH$^1$N$^1$)x3x1*. As the errors are respectively 21.5% ± 0.05% and 25.4± 0.2%, we can deduce that the generated error is location-dependent. In addition, we have shown that the use of exogenous variables improved the *MLP* prediction mainly during winter and autumn (gain of 0.7 point). Similar results are obtained with the *PV* energy forecasting [44].

The main conclusions for this d+1 horizon can be resumed as following:

- without the use of exogenous variables, *MLP* are equivalent to *ARMA* (*nRMSE*~22%);

- for cloudy months (winter and autumn), the use of exogenous variables improves the quality of the prediction (gain of 0.7 point);

- make the *TS* stationary with $k^*$, or if possible *CSI$^*$* is appropriate (gain of 1.5 points);

- persistence is an interesting naive predictor, which gives very good results in spring and summer (*nRMSE*=26.1%);

- the prediction methodology is applicable in the global radiation case and *PV* energy case.

This first study on the daily horizon allows us understanding how to use the *MLP* and other predictors studied. We showed that tested predictors like Markov, Bayes and k-NN are relatively equal in terms of prediction. The details of this comparison are given in [44]. These predictors proved to be much less suitable for predicting global radiation as *ARMA* or *MLP*. With this result, we decide in the following to not use the Markov, Bayes and k-NN estimators. For the naive estimator, only the persistence will be used for its ease of use and good results, especially on sunny days (*nRMSE*= 19% in May and June).

## 4.2. Hourly case

For this horizon the *CSI\** approach simplifies the *MLP* architecture: one endogenous input and a maximum of 8 hidden neurons for the five *TS* studied. But this does not improve the prediction error, so in the following, the stationary mode will not use the periodic coefficients. Performing the same study in the case of *ARMA* predictions, *CSI$^*$* and *CSI* stationarization give similar results. Henceforth, we will therefore use the *CSI* with these predictors. Note that the clearness index generates less efficient results [45]. The Table 4 presents the comparison of seasonal *nRMSE* related to estimators for global radiation for the five cities. For predictions with *MLP*, we study the case with only the endogenous variables (*MLP*endo) and the combining of endogenous and exogenous variables (*MLP*endo-exo).

| Models | Anual | Winter | Spring | Summer | Autum |
|---|---|---|---|---|---|
| Persistence | 36.0±2.0 | 46.4±5.1 | 36.3±3.4 | 30.0±3.2 | 41.5±3.3 |
| *ARMA* | **16.4±0.7** | 22.2±1.8 | **15.9±1.2** | **14.1±0.9** | 19.4±0.6 |
| *MLP*endo | 17.0±1.2 | 20.3±2.0 | 17.0±2.3 | 15.6±1.2 | 18.1±1.1 |
| *MLP*endo-exo | 16.8±1.3 | **20.2±2.1** | 16.9±2.2 | 15.5±1.4 | **17.5±14** |

Table4: performance comparison (*nRMSE* and confidence interval in %) between different studied models (average on the five cities). Bold characters represent the lowest values

In summer, the interest of methods like *MLP*endo and *MLP*endo-exo is minimal. This is undoubtedly due to the low probability of occurrence of clouds during this period. A linear process like *ARMA* seems best suited. We can probably conclude that use of *MLP* with endogenous and exogenous variables is interesting when the cloud cover is intense (mainly in autumn and winter). In [45] we have shown that the predictors hybridation (*ARMA* and *MLP*endoexo) increases the quality of predictions. The method used is based on the following selection rule:

$$if |\varepsilon^{AR}(t)| \leq |\varepsilon^{PMC}(t)| \ then \ \hat{X}(t+1) = \hat{X}^{AR}(t+1) \ else \ \hat{X}(t+1) = \hat{X}^{PMC}(t+1)$$

Equation 3

The Figure 1 shows the average gain (computed on the five cities) of *nRMSE* obtained by the hybrid method compared to the better *MLP* (grey bars) and the better *ARMA* (dashed bars). The gain is positive when the hybridization is better than traditional methods.

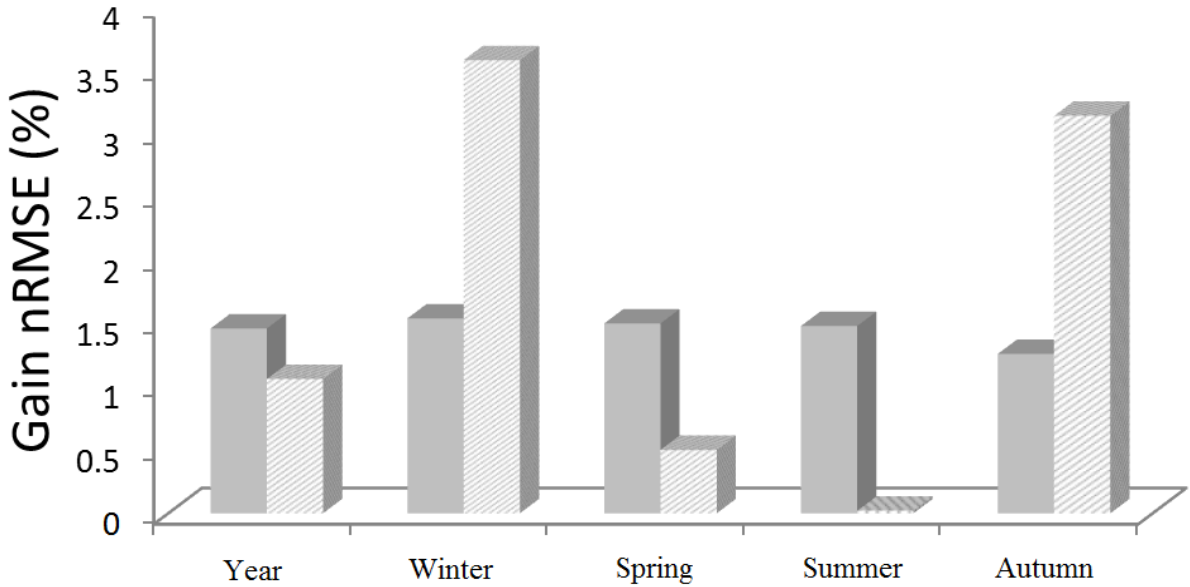

Figure 1: mean gain related to the hybrid model compared to the models *MLP* (grey bars) and *ARMA* (dashed bars)

The maximum gain is observed in winter (3.8 ± 0.8% better than the *ARMA* model) and the minimum is in summer, when the hybrid method is as interesting as the *ARMA* method (gain of 0.02 ± 0.5%). For all sites, it is clear that the hybrid model approximates correctly the global radiation [45]. In previous study[45] we have shown that exogenous data (meteorological measures) can be replaced by estimation of analytic models like the numerical weather prediction model ALADIN [45]. In this context, the results generated by hybrid *MLP/ARMA*, ALADIN and *CSI*[*] should be different (see the table 5).

|  | *ANN*ual | Winter | Spring | Summer | Autumn |
|---|---|---|---|---|---|
| **Ajaccio** | 14.9(25.1) | 19.4(34.7) | 15.5(25.2) | 11.0(21.4) | 17.0(33.9) |
| **Bastia** | 16.5(27.1) | 19.5(35.0) | 17.5(27.1) | 13.2(22.6) | 17.9(34.4) |
| **Montpellier** | 14.7(26.9) | 15.7(32.6) | 15.2(25.9) | 13.4(24.6) | 15.5(33.2) |
| **Marseille** | 13.4(25.3) | 16.6(32.9) | 14.8(25.3) | 9.3(20.0) | 13.8(32.3) |
| **Nice** | 15.3(26.4) | 16.6(32.1) | 15.3(24.5) | 10.3(21.1) | 26.2(37.1) |

Table5: prediction error (*nRMSE* %) for the hybrid model *ARMA*, *MLP*, *ALADIN*, *CSI*[*], the persistence results are presented between parenthesis.

This hybrid model is very interesting: the 10% threshold has been crossed in Marseille. Although summer is the season where the hybrid methodology is the less interesting, all seasons and cities benefit from this hybridization model. We can note that *MLP* and *ARMA* are very effective alone in summer period. To resume, use of the hybrid method reduces the error by 11% compared to the prediction done by persistence (mean on the five cities).

In summary, the fact to make stationary the global radiation *TS* reduces the error by $0.5 \pm 0.1\%$ for the five locations studied. The use of ALADIN and of hybridization models shows a real potential and a strong interest. This step allows to increase significantly the quality of the prediction (gain close to 3.5 points). In the end, if we compare this approach with a simple prediction such as persistence, there is a reduction of the prediction error of more than 11%.

The methodology of prediction based on *CSI*, ALADIN *MLP* and *ARMA* is certainly complicated to implement, but gives results far superior to those from other tested techniques. We note that for this horizon, the *CSI* must be used to overcome seasonal variations. In addition, the use of exogenous variables is an added value to the modeling. Forecasts of meteorological variables from ALADIN model offer prediction accuracy. However, the use of meteorological measurements gives also good results, although less efficient. Finally, the combination of all the improvements that we recently proposed amplifies the quality of the prediction.

### 4.3. 24-hours ahead case

This new horizon studied is the prediction for the next day hour by hour [10,51] of the global radiation profile. Unlike hourly, daily or monthly horizons, this horizon is little discussed in the

literature.We may mention the work of MellitandPavan[27] which propose to useas input of theprediction tool (*MLP*) thedaily mean valuesof solar radiationand temperature, and the dayof the considered month. To satisfythispredictionhorizon, we have considered approaches basedon the use of*MLP*, followingconclusionspresentedearlier in this paper.As a first step, we focus on the endogenous case, and then we will introduce exogenous parameters. The predictor is a *MLP*like in the previous case, but with multiple outputs (one by hours).Measurementsare chronologically positionedinthe input vectorof *MLP*. We choose to compare the *MLP* resultswith those obtained bymethods ofpersistenceand*ARMA*. The last method we have tested is basedon multiple*ARMA* models which each are dedicated to one particular hour. Note thatall thesemethodsare compatible withthe use of theclearness index($k$and$k^*$) and the clear sky index (*CSI*and*CSI$^*$*). Moreover, in h+1 and d+1 horizons, the seasonal adjustments didnot show strong superiority. For these reasons, in the next manipulations only *k* and *CSI* will be considered. The goal is to find a relatively simple and generalizable methodology taken care of not draw conclusions about data snooping.Results are shown in the table 6.

| Type | | *ANN*ual | Winter | Spring | Summer | Autumn |
|---|---|---|---|---|---|---|
| Persistence | | 35.1 | 54.8 | 35.2 | 28.0 | 40.4 |
| ARMA | k | 29.1 | 44.6 | 29.2 | 24.0 | 33.2 |
| | CSI | 28.6 | 44.2 | 28.6 | 23.1 | 32.8 |
| MLP | k | 27.9 | 44.2 | **27.9** | 22.2 | 32.7 |
| | CSI | **27.8** | **42.8** | 28.4 | **22.0** | **31.3** |

Table 6:*nRMSE*(%) of predictions realized with the *MLP*. Boldcharacters represent the best results.

We note thatsophisticated approachesas*ARMA*or*MLP*largelyoutperformnaive modelespecially in winter. Note alsothat the bestpredictions areobtained withthe use of theclear skyindex (*CSI*).Contrary to the previous case (h+1 case),the *MLP* is systematically better than *ARMA* model. The interest of a hybrid approach seems for this reason not relevant. However,it is possible tointegrate exogenous inputs. After several trials, we found that the more interesting data are the hourly pressure and cloudiness of the last day, and the daily average nebulosity of the two last days.The contribution of these variables is presented in the Table 7(only the *CSI* methodology is shown because more interesting).

| Type | *ANN*ual | Winter | Spring | Summer | Autumn |
|---|---|---|---|---|---|
| Persistence | 35.1 | 54.8 | 35.2 | 28.0 | 40.4 |
| *ARMA* | 28.6 | 44.2 | 28.6 | 23.1 | 32.8 |
| *MLPendo* | 27.8 | 42.8 | 28.4 | 22.0 | **31.3** |
| *MLPexo* | **27.3** | **42.4** | **27.8** | **21.7** | **31.3** |

Table7: impact of exogenous variables on the prediction quality (*nRMSE* %). In bold the best results.

For the h+24 horizon the contribution of exogenous variables is less explicit that for previous case studied. These kind of deep horizons (≥24 h) modify approach to consider. Thus, this type of prediction is particularly difficult to realize. Search the smoothness of a 24-hours ahead prediction depends on too many parameters to expect to get the same level of results as for horizons h+1 or d+1. We can conclude that it is valuable to make stationary data (*nRMSE* gain close to 0.5 point). To do this the use of clear sky index is preferable, even if the clearness index gives results almost similar. The *CSI* allows a *nRMSE* gain of 0.5 point for *ARMA* and 0.1 point for *MLP* related to the *k* use. The classical approach involving a single *MLP* with multiple outputs is recommended: *nRMSE* gain of 0.6 point for *k* index and 0.4 point for *CSI* index related to a *MLP* committee like described in the *ARMA* case. In the present state of our knowledge, the ratio between performance and complexity induces, to not use exogenous variables (maximal *nRMSE* gain of 0.6 point in Winter).

### 4.4. Five minutes case

By its nature this prediction horizon is completely different from what we have studied so far. The originality of this case is the sampling frequency of measurement that is less than the dynamics of cloud occurrence. Thus, in 5 minutes the sky has a high probability of remain identical. Data are available on the *PV* wall of Vignola laboratory [44]. They cover the period from March 2009 to September 2010. The installation allows identifying three separate areas: 0 °, 45° SE and 45 °SW tilted at 80° relative to the ground surface.

| *Orientation / Type* | | Total | May | June | July | August |
|---|---|---|---|---|---|---|
| SW | MLP | **21.4** | **31.4** | 20.7 | **14.2** | **19.5** |
| | MLP + k | 22.5 | 32.3 | **20.1** | 15.4 | 19.6 |
| | MLP + CSI | 22.2 | 31.9 | 21.1 | 16.3 | 20.0 |
| | Persistence | 21.8 | 32.3 | 20.9 | 14.4 | 19.6 |
| S | MLP | **20.2** | **28.0** | 22.6 | **13.5** | **16.5** |
| | MLP + k | 21.7 | 29.6 | 23.7 | 14.8 | 18.4 |
| | MLP + CSI | 21.9 | 29.7 | 25.5 | 17.4 | 19.5 |
| | Persistence | 20.8 | 28.8 | 23.2 | 13.8 | 17.1 |
| SE | MLP | **23.2** | **31.8** | **26.5** | **14.6** | **20.6** |
| | MLP + k | 24.2 | 32.6 | 27.6 | 15.1 | 21.7 |
| | MLP + CSI | 25.6 | 33.3 | 28.1 | 17.8 | 23.8 |
| | Persistence | 24.5 | 33.3 | 27.9 | 14.8 | 22.0 |

Table8: Stationary process impact on the error of prediction (*nRMSE* in %)

The Table 8shows the impact of the stationary process. Unlike in the daily and hourly case this studydoes not allow concluding that theuse of *CSI* and *k*arejustified.For this tilt and orientation, the theoretical models are limited. In these configurations the solar shield complicated the phenomena. For this reason, *CSI,k, CSI$^*$* and *k$^*$*are not used in the following (only raw data).

In fact, in the raw global radiation *TS*, output of*MLP*correspondsto animproved persistence. As the prediction seems to be a persistence(delay of 5 min), weights related to the first lag are important and other are close to zero.

Simpler tools, accessible with*MLP*could improvethe prediction results. Indeed, the*MLP* can alonechooseits ownstationarity, usingas inputtime indexes,which will enable itto establish aregression on thetimeof the periodic phenomenon.The twotime indicesused arerelated tohour of the dayandday of the year. The transfer functionin the hidden layerwhichgives the best resultsisthe Gaussian function.The use oftime indexgeneratesanadded value tothe quality of theprediction. Results aresystematically improvedby this tool:*nRMSE*is reducedby0.7 point for the SWand S orientations and 0.1% in the SE case. The average gain isgreater than0.5 point, ensuring a realadvantage inusing thisstationarization mode. Table 9 shows the results obtained.

|    | *MLP* | *MLP*+time index | persistence |
|----|-------|------------------|-------------|
| SE | 23.2  | 23.1             | 24.5        |
| S  | 20.2  | 19.5             | 20.8        |
| SW | 21.4  | 20.7             | 21.8        |

Table9: prediction error (*nRMSE*in %) related to the *MLP* and the time index methodology

Note that for this horizon, the use of *ARMA* is not relevant because the optimization led us to use an simple *AR*(1) where the regression coefficient of lag 1 is close to 1. This kind of model is in fact persistence. Like *MLP* is systematically better than persistence, the hybridization of models is not justified. Moreover, the use of exogenous data does not provide benefit for the prediction. Furthermorethere are veryfew measurementswith asamplingnear5 minutes. This kind of prediction process is very complicate to construct. In brief, we have seenin this section thatmethodsused to make stationary the *TS* are not available for this horizon (*nRMSE*increasedby 1 point). It is more appropriateto usethe raw seriesand not theclear skyor clearness index, but the use of time index is interesting to takeinto account the seasonality. We may also notethat the*MLP*-based methodology improves outcomes (*nRMSE*improved to more than 1 point)compared to a simpler approach based on persistence.

# 5. Conclusion

In all bibliographic items related to the estimation of global radiation, we find that the errors associated with predictions (monthly, daily, hourly and minute) differ from sites and from authors. Methodologies of predictions are usually so different that they are difficult to compare. In addition, the estimations errors are heterogeneous: prediction error on certain days or sampled over an extended period, test on the cumulative predictions, use of non-standard error parameters, etc. To overcome all these features we present here is a methodology of comparison of different predictors developed and tested to propose a hierarchy. Only the *TS* approach is studied, other weather models using numerical weather prediction models or satellite images are not considered. For horizons d+1 and h+1, our results are partly consistent with the literature. Indeed, *MLP* are adapted and used to make predictions of global radiation with an acceptable error [52] and are also applicable to mountainous areas [53]. Regarding prioritization of *ARMA* and *MLP*, the results shown here are different from traditional bibliographic results [26,54,55]. In fact, without stationarity we do not think it is easy to differentiate between *ARMA* and *MLP*. Moreover, while *ANN* by its non-linear nature is effective to predict cloudy days, *ARMA* techniques are more dedicated to sunny days without cloud occurrences. However, we agree Berhangh et al. [37] with the fact that the use of exogenous variables improves the results of *MLP*. As in the literature, we found that the relevant approaches in the case of the prediction of radiation were equally in the case of the prediction of *PV* power [26,56]. Although it is not routinely used in the literature, we believe that persistence can correctly judge the validity of complex technical and we chose as naive predictor. In literature, clear sky model and seasonal adjustments based on periodic coefficients have not often been used with the prediction of global radiation. The views of the results presented here, their investigation looks promising. Finally, for horizons h+24 and m+5, there are still too few studies using the *MLP*. However as Mellit and Pavan [27] and Chaabene and Ben Ammar [57] we believe and have shown that the *MLP* were adapted to these situations. In addition, our approach with the use of time index appears to be efficient. In summary, our results are complementary and improve the existing prediction techniques with innovative tools (stationarity, *NWP* combination, *MLP* and *ARMA* hybridization, multivariate analysis, time index, etc.).

Through this work, we have identified some methodologies for the prediction horizon of global radiation. We can conclude that these two types of predictions are relatively equal in the methodology to implement. In Table 10 are listed and summarized *TS* based methods we recommend for different prediction horizons.

| Horizons | stationarity | Exogenous data | Required predictors | difficulty | *nRMSE* |
| --- | --- | --- | --- | --- | --- |
| d+1 | *CSI\** | Measures: *Su.N.RH* | *MLP (>ARMA>pers)* | ++ | 23.4% |
| h+1 | *CSI* | NWP: *N. P. RP* | *Hybrid_MLP+ARMA (>MLP>ARMA>pers)* | +++ | 14.9% |
| h+24 | *k* | - | *MLP multi-outputs (>multiMLP>ARMA>pers)* | + | 27.3% |
| m+5 | *Time index* | - | *MLP (>ARMA>pers)* | + | 20.2% |

Table10: summary of the resultspresented in this paper

In view of the previous manipulations, we note that the results can be completely different depending on the time horizon.For this reason, we must pay attention to themethods usedand the expected results. What should besought isa simple methodto implement, cost effective and workablein several locations:the selection ofdata andmodel parametersmust bechosenparsimoniously. To conclude this paper, we believe that the establishment of a benchmark in the areas of renewable energy would allow the community to better share, understand and interpret the results: same data, comparisons of models using the same tools RMSE, *nRMSE*, IC95%, etc. The recent European COST (Cooperation in Science and Technology)initiative called WIRE(Weather Intelligence for Renewable Energies)[3]seems to follow this idea and should be encouraged.

## Acknowledgement

Thanks to an agreement with Météo-France, which is the French meteorological organization, we had the opportunity to freely access to some of their forecasts and measures.

---

[3] http://www.cost.eu/domains_actions/essem/Actions/ES1002

# List of Tables

# List of Figure